\documentclass[aps,prl,twocolumn,showpacs,superscriptaddress]{revtex4}

\newcommand{\ket}[1]{|#1\rangle}
\newcommand{\bra}[1]{\langle#1|}
\newtheorem{theorem}{Theorem}

\usepackage{amssymb}
\usepackage{amsmath}
\usepackage[sort&compress]{natbib}
\usepackage[dvips]{graphicx}

\begin{document}


\title{Quantum coin tossing 
and bit-string generation
in the presence of noise}
\author{Jonathan Barrett}
\email{jbarrett@ulb.ac.be}
\affiliation{Service de Physique Th\'{e}orique, Universit\'{e} Libre
de Bruxelles, CP 225, Bvd. du Triomphe, 1050 Bruxelles, Belgium} 
\affiliation{Th\'{e}orie de l'Information et des Communications, CP
165/59, Universit\'{e} Libre de Bruxelles, Av. F. D. Roosevelt 50,
1050 Bruxelles, Belgium} 
\author{Serge Massar}
\email{smassar@ulb.ac.be}
\affiliation{Service de Physique Th\'{e}orique, Universit\'{e} Libre
de Bruxelles, CP 225, Bvd. du Triomphe, 1050 Bruxelles, Belgium} 
\affiliation{Th\'{e}orie de l'Information et des Communications, CP
165/59, Universit\'{e} Libre de Bruxelles, Av. F. D. Roosevelt 50,
1050 Bruxelles, Belgium} 

\begin{abstract}

We discuss the security implications of noise for quantum coin tossing
protocols. We find that if quantum error correction can be used, so that noise levels can be made arbitrarily small, then reasonable security conditions for coin tossing can be framed so that results from the noiseless case will continue to hold. If, however, error correction is not available (as is the case with present day technology), and significant noise is present, then tossing a single coin becomes problematic. In this case, we are led
to consider random $n$-bit string generation in the presence of noise,
rather than single-shot coin tossing. We introduce precise security
criteria for $n$-bit string generation and describe an explicit
protocol that could be implemented with present day technology. In general, a cheater can exploit noise in order to bias coins to their advantage. We derive explicit upper bounds on the average bias achievable by a cheater for given noise levels.  

\end{abstract}

\pacs{03.67.Dd, 03.67.Pp, 05.40.Ca}

\maketitle

The cryptographic task of coin tossing was first introduced by Blum
\cite{blum1982}. Briefly, the idea is that two separated, mistrustful
parties wish to generate a random bit and be sure that the other party
cannot have biased the bit by more than a certain amount. Secure coin
tossing is known to be impossible classically, unless either
computational assumptions or special relativistic considerations
\cite{kent1999} are invoked. Upon using a quantum communication
channel, however, it is possible to achieve 
levels of security that are
impossible classically. Various results concerning the security of
quantum coin tossing under different assumptions, and its relationship
to other cryptographic tasks (in particular, bit commitment), have
been obtained
\cite{mayers1997,lochau1998,kent1999,aharonovta-shmavaziraniyao2000,spekkensrudolphdegrees2002,kent2001,spekkensrudolphcheatsensitive2002,ambainisnewprotocol2002,ambainislowerbound2002,doscherkeyl2002,kerenidisnayak2002,nayakshor2002,kent2002,kitaev}.  


It would 
be highly desirable to implement quantum coin
tossing experimentally. Long distance quantum communication is indeed possible with
present technology \cite{ribordy1998}, 
and it may be possible to adapt such techniques
to coin tossing. In real life situations, however, state
preparation, communication channels and measurements are all
imperfect, while all the above results refer to the ideal
situation where no noise is 
present. As the example of quantum key distribution
illustrates, a 
large amount of theoretical work must be
carried out before an idealized quantum communication protocol can be
implemented experimentally with the effect of all experimental
imperfections taken into account. The present work initiates this line
of investigation in the case of quantum coin tossing. 

In particular, we shall argue that if quantum error correction can be used, then noise can in principle be made arbitrarily small, and one can frame security definitions such that results holding in the noiseless case still apply. On the other hand, if techniques such as quantum error correction are not available (as is the case with present day technology), then there will be a significant level of noise that cannot be reduced. In this case, tossing a single coin is problematic, in the sense that it is not possible to frame sensible security conditions that can actually be satisfied. But tossing a string of coins such that the
average bias is bounded is possible. We obtain detailed results
concerning the relation between the amount of noise and this average
bias. 

We begin with some definitions that have been introduced for noiseless coin tossing (see, e.g., Ref.~\cite{spekkensrudolphcheatsensitive2002}). A coin tossing protocol involves a sequence of rounds of communication, at the end of which either a bit $x$ is produced, whose value is agreed on by both parties, or one party or the other aborts, in which case we write symbolically $x=\infty$. We denote by $\mathrm{Pr}^{S_A,S_B}(x=c)$ the probability that $x=c$, assuming that Alice follows a strategy $S_A$ and Bob $S_B$. We denote Alice's honest strategy (i.e., that defined by the protocol) as $H_A$ and Bob's as $H_B$.
A protocol is \emph{correct} if, for $c=0,1$
\begin{equation}
\mathrm{Pr}^{H_A,H_B}(x=c)=1/2.
\end{equation}
Security conditions are written as
\begin{eqnarray}
\forall \, S_A \ \mathrm{Pr}^{S_A,H_B}(x=0)&\leq& 1/2 + \epsilon_A \label{weakalicecondition}\\
\forall \, S_A \ \mathrm{Pr}^{S_A,H_B}(x=1)&\leq& 1/2 + \epsilon_A \label{strongalicecondition}\\
\forall \, S_B \ \mathrm{Pr}^{H_A,S_B}(x=0)&\leq& 1/2 + \epsilon_B \label{strongbobcondition}\\
\forall \, S_B \ \mathrm{Pr}^{H_A,S_B}(x=1)&\leq& 1/2 + \epsilon_B \label{weakbobcondition}.
\end{eqnarray}
(These conditions define the task known as \emph{strong coin tossing}. A weaker task can be defined by imposing only (\ref{weakalicecondition}) and (\ref{weakbobcondition}). This is known as \emph{weak coin tossing}. In this work we are concerned only with strong coin tossing. With this understood, we shall simply call it coin tossing.) A protocol is \emph{perfectly secure} iff $\epsilon_A=\epsilon_B=0$. A protocol is \emph{arbitrarily secure} iff $\epsilon_A$ and $\epsilon_B$ can be made arbitrarily small as some parameter associated with the protocol increases. A protocol is \emph{partially secure} iff $\epsilon_A,\epsilon_B<1/2$. Perfectly secure coin tossing is shown to be impossible in Ref.~\cite{lochau1998}. More recently, Kitaev has shown \cite{kitaev} that for any possible protocol, either $\epsilon_A$ or $\epsilon_B$ $\geq 1/\sqrt{2}-1/2$. This result implies that arbitrarily secure strong coin tossing is impossible. At present, the best protocol for strong coin tossing is due to Ambainis \cite{ambainisnewprotocol2002} and achieves $\epsilon_A=\epsilon_B=1/4$ (close to Kitaev's lower bound).  

In order to discuss what happens when noise is present, we first describe a very simple protocol for coin tossing in the
absence of noise (it is similar to the protocol for quantum gambling developed in Ref.~\cite{hwangahnhwang2001}). The protocol is for strong coin tossing and is
partially secure. It is not as good as that of Ambainis, but is
illustrative. We shall then consider how it is affected by noise. The protocol is as follows.

{\bf i)}
Alice generates a random bit $b\in\{0,1\}$, and prepares a quantum
state $\ket{\phi_b}$, where $0<|\langle\phi_0|\phi_1\rangle|^2<1$. We
write $|\langle\phi_0|\phi_1\rangle|^2\equiv \cos^2\theta$. Alice
sends $\ket{\phi_b}$ to Bob. 

{\bf ii)}
Bob generates a random bit $b'$ and sends it to Alice.

{\bf iii)}
Alice sends $b$ to Bob.

{\bf iv)}
Bob measures the quantum state in a basis containing $\ket{\phi_b}$ to
check that Alice is not cheating. He aborts if he gets an outcome
different from $\ket{\phi_b}$. In this case we write the outcome of
the coin toss as $x=\infty$. Otherwise, the outcome of the coin toss
is $x=b\oplus b'$. 

It is easy to show that Alice's best cheating strategy is to send a
state $\ket{\chi}$ such that
$|\langle\chi|\phi_0\rangle|^2=|\langle\chi|\phi_1\rangle|^2=\cos^2\theta/2$,
and then to declare $b$ as she requires. (In particular, it is never to her advantage to declare the opposite value of $b$, and this implies that strategies that involve her entangling two systems and sending one to Bob cannot help.) Bob's best cheating strategy
is to measure the quantum state as soon as he receives it, in order to
determine as well as possible whether it is $\ket{\phi_0}$ or
$\ket{\phi_1}$ \cite{aharonovta-shmavaziraniyao2000}. We get that 
for $c=0,1$,
\begin{eqnarray}
\forall \, S_A \ \mathrm{Pr}^{S_A,H_B}(x=c)&\leq& 1/2+(1/2)\cos\theta\nonumber\\
\forall \, S_B \ \mathrm{Pr}^{H_A,S_B}(x=c)&\leq& 1/2+(1/2)\sin\theta.
\end{eqnarray} 
Note that the protocol is correct.

How are this protocol, and the corresponding security considerations,
affected if it is assumed that noise is present? We discuss mainly the
case in which the quantum channel separating Alice and Bob is noisy,
but in which all other devices are perfect. 
Other types of
experimental imperfections that could be considered include, for
example, noise in state preparation and measurement, or the limited
efficiency of detectors. 
The arguments we use in the case of noisy channels can easily be
adapted to these other situations.
In general, the channel can be described by  
a linear, completely positive, trace-preserving map ${\cal E}$.
Thus Bob will receive the noisy state $\rho_a = {\cal E} (|\psi_a \rangle)$.
For long distance quantum communication the principal type of noise
will be losses. Losses can in principle be included in the
form of ${\cal E}$,  but it may be convenient to consider them
separately.
An honest party must assume that the other, potentially dishonest,
party can control everything outside the honest party's laboratory. In
particular, this means that a cheater can replace the noisy channel
with a noiseless one, introducing noise only as and when he or she
wants. 

The presence of noise on the face of it affords several simple
cheating strategies. We list a few. 

{\bf i)} 
If the channel is lossy, and the rules specify that if Bob's detector
does not fire then he simply has to accept whatever value of $b$ Alice
declares, then Alice may cheat simply by not sending any quantum state
at all. Bob will think that the state was lost in the channel. Alice
will declare whatever value of $b$ she wants, winning with certainty.  

{\bf ii)} 
Suppose alternately that the rules specify that if Bob receives no
state, he is allowed to declare ``no fire,'' and the protocol
recommences. Then a cheating Bob may replace the noisy, lossy channel
with a noiseless one, so that the states he must discriminate are
pure. He then (using a perfect detector) performs a conclusive
measurement on the quantum state, with outcomes ``definitely
$\ket{\phi_0}$,'' ``definitely $\ket{\phi_1}$,'' and ``inconclusive''
\cite{peres}. If the ``inconclusive'' result is obtained, then Bob
declares ``no fire,'' and the protocol repeats. The protocol will keep
repeating until a run occurs on which Bob finds out the value of $b$
with certainty. He can then obtain the outcome he wants with certainty
by choosing $b'$ appropriately.  

{\bf iii)}
In general, both $\rho_0$ and $\rho_1$ will have support on the entire
Hilbert space. This means that a cheating Alice can send
$\ket{\phi_0}$ to Bob, and then declare that $b=0$ or $b=1$ as she
needs. Even in the event that Alice declares $b=1$, there will be no
measurement result of Bob's that tells him for certain that Alice is
cheating.  

In addition, if the channel is noisy and if a certain outcome of Bob's leads him to
abort the protocol, then there is a nonzero probability that he may
obtain this result even when Alice is honest. In this case, the
protocol is no longer correct, because there is a nonzero probability
that a player aborts, even when both are honest. 

It should be clear that most of these remarks will apply in some form
or other to any conceivable coin tossing protocol. If correctness is redefined as
\begin{eqnarray} 
\mathrm{Pr}^{H_A,H_B}(x=c) &=& (1-\delta)/2 \nonumber\\
\mathrm{Pr}^{H_A H_B}(x=\infty) &=& \delta \label{correctnesswithnoise},
\end{eqnarray}   
for $c=0,1$, where $\delta$ must tend to zero as some parameter associated with the protocol increases, and if similar modifications are made to the security conditions, then protocols can still be made (at least partially) secure, as long as error correction is available. In this work, however, we are interested in the case in which there is no technique for making noise arbitrarily small. Protocols for
tossing a single coin will then either allow one
of the parties to bias the coin completely, or will abort with unacceptably high probability even though both parties are honest.
They are therefore of limited interest.
For these reasons we consider instead a slightly different
scenario, in which Alice and Bob want to generate a random $n$-bit
string. 


The basic reason why random $n$-bit string generation is easier to
implement in the presence of noise is that, since the quantum channel
is used many times, one can test that the average noise level is as
expected. This is to be contrasted with a single use of the channel,
which only provides very partial information about the noise
present. Furthermore the probability that the protocol aborts 
when both parties are honest decreases exponentially with $n$. This is
due to the fact that the channel is used many times, which makes large
fluctuations about the expected noise highly improbable. 
On the other hand, because the noise is only constrained on average, a
cheater can determine any bit of the string with certainty. This
implies that one cannot extract from the bit string a single unbiased
bit (for instance by taking the parity of the bit string). But the
average bias of the bits can be bounded \footnote{It may be instructive to contrast this situation with that of noisy key distribution. In the case of noisy key distribution, if quantum error correction and entanglement distillation are not available, the raw key will only be partially secure. However, there is still the possibility of extracting an arbitrarily secure key via postprocessing of the raw key data. In the case of bit-string generation, on the other hand, there is no possibility of ``distilling'' an arbitrarily secure single coin toss from the partially secure bit string.}.

The problem of generating a random $n$-bit string in the absence of
noise has previously been considered by Kent \cite{kent2002}, who
showed that it is not a straightforward extension of the problem of
generating a single random bit. 
We now introduce some precise security criteria for $n$-bit string generation. 
The output of a protocol for $n$-bit string generation is $\vec{x}$,
where either $\vec{x}$ is an $n$-bit string, or one or the other party
aborts, in which case we write symbolically $\vec{x}=\infty$.  

One security condition that one could imagine 
for $n$-bit string generation is that
each bit of the string has small bias. This could be expressed as
\begin{eqnarray}\label{bitwisesecurity}
\forall {S_A} \forall i\  
\mathrm{Pr}^{S_A, H_B}(x_i = c) \leq 1/2 + \epsilon_A, 
\end{eqnarray}
for $c=0,1$, along with a similar condition for Bob.
Here, $x_i$ is the $i$th bit of $\vec x$. As we have argued above, this type of
security is not achievable in the presence of noise.

A weaker security condition is that on average the bias of the bits is
small. We express this as 
\begin{eqnarray}\label{foru}
\forall S_A  \forall \vec c\ 
\frac{1}{n} \sum_i
\mathrm{Pr}^{S_A, H_B}(x_i = c_i) \leq 1/2 + \epsilon_A,
\end{eqnarray}
along with a similar condition for Bob. We have that $x_i$, $c_i$ are
the $i$th bits of $\vec{x}$,$\vec{c}$, with $\vec{c}$ an arbitrary
$n$-bit string describing a possible result of the coin tosses. 

Other security conditions are possible. Satisfaction of condition (\ref{bitwisesecurity}) or (\ref{foru}) is compatible with a cheater fixing things so that the outcome is either $00\ldots 0$ with probability $1/2$ or $11\ldots 1$, with probability $1/2$. It would be desirable to have a security condition expressing the fact that the
entropy of the bit string is large, satisfaction of which would rule out such cheating.
In the remainder of this article, however, we will restrict ourselves to
the security condition expressed by Eq.~(\ref{foru}). In general, $\epsilon_A$ and $\epsilon_B$ will depend on $n$. For simplicity we shall only be interested in the values of these quantities in the limit of large $n$.

We define correctness by
\begin{eqnarray} 
\forall \vec{c} \quad \mathrm{Pr}^{H_A,H_B}(\vec{x}=\vec{c}) &=& (1 - \delta)/2^n \nonumber\\
\mathrm{Pr}^{H_A,H_B}(\vec{x}=\infty)&=&\delta,
\end{eqnarray}
where $\delta$ must tend to zero as $n$ becomes large.

We now introduce a protocol for random $n$-bit string generation that is
adapted from the simple coin tossing protocol above. 
Then we will consider the security of this protocol in the presence of
noise.

{\bf 1) For $i=1$ to $n$:}

{\bf i)}
Alice generates a random bit $b_i\in\{0,1\}$, and prepares a quantum
state $\ket{\phi_{b_i}}$, where
$0<|\langle\phi_0|\phi_1\rangle|^2<1$. Alice sends $\ket{\phi_{b_i}}$
to Bob. 

{\bf ii)}
Bob generates a random bit $b_i'$ and sends it to Alice.

{\bf iii)}
Alice sends $b_i$ to Bob.

{\bf iv)}
Bob measures the quantum state in a random basis. If his detector
fails, this is considered as a null outcome. 

{\bf Next $i$.}

{\bf 2)}
Bob uses his measurement statistics
to estimate $\rho_0$, the average state he received when Alice declared $b_i=0$, and $\rho_1$, the average state he received when Alice declared $b_i=1$. If either of the fidelities, $\bra{\phi_0}\rho_0\ket{\phi_0}$ and $\bra{\phi_1}\rho_1\ket{\phi_1}$, is less than $1-\gamma$ (where $0\leq \gamma \leq 1$ is decided in advance), then Bob aborts the protocol. Otherwise the output of the protocol is an $n$-bit string with $x_i=b_i\oplus b_i'$. 

This protocol is essentially the simple protocol above repeated $n$
times, with two modifications. First, Bob measures each time in a
random basis - he is performing a sort of state estimation in order that he can bound any potential cheating by Alice. Second, an honest Bob does not have the
option of aborting until the end of the protocol when he has collected
all his statistics. It is easy to see that if the fidelity of the whole process of state preparation, transmission and measurement is $F>1-\gamma$, then the protocol is correct.

We investigate available cheating strategies for this protocol,
assuming that the only noise is noise in the quantum channel,
described by ${\cal E}$, and that either there are no losses or they
are included in the form of ${\cal E}$. We can consider two cases.  
In the first, a cheater's actions on different runs (that is,
different values of $i$ in the protocol above) are uncorrelated. Thus,
we simply need to consider one strategy, perhaps involving random
choices, that is repeated for each run. In the second, a cheater's
actions on different runs may be correlated, and may even involve
entanglement across the different runs \footnote{In some ways, the
assumption of uncorrelated cheating strategies may be likened to the
assumption of incoherent attacks that is common in analysis of quantum
key distribution.}. In this paper, we will restrict ourselves to uncorrelated cheating.
Elsewhere, we show that this protocol in fact satisfies an entropic security condition, even when correlated or entangled attacks are considered \cite{barrettmassar}. 

A cheating Bob is easiest to deal with. His best strategy is to
replace the noisy channel with a noiseless one, thus ensuring that the
states he receives are $\ket{\phi_0}$ and $\ket{\phi_1}$.  
He can measure each state as soon as he receives it, in order to
determine as well as possible the identity of the state. He can then
choose $b_i'$ appropriately. As mentioned before, this gives
$\epsilon_B=1/2\,\sin\theta$, where
$|\langle\phi_0|\phi_1\rangle|^2=\cos^2\theta$. 

The most general strategy for a cheating Alice is to prepare a pure
state $\ket{\psi}_{AB}$, and send the $B$ subsystem to Bob via a
noiseless channel. (In general, of course, Alice may prepare an
overall mixed state, perhaps resulting from a probabilistic mixture of
pure states. We lose no generality, however, by supposing that Alice
prepares a pure state, as Alice can always introduce an extra ancilla
such that $\ket{\psi}_{AB}$ is a purification of the mixed state.) We
denote the reduced density matrix for Bob's subsystem by $\rho_B$. 
Alice then waits for the bit $b_i'$. 
The value of $b_i'$ and the outcome of the coin toss that she wants
determine together the value of $b_i$ that Alice wants to declare.  
If Alice wants to declare $b_i=0$, then she performs a two-outcome
positive operator-valued (POV) measurement $M_0$ on the $A$
subsystem. Denoting the outcomes $M_{00}$ and $M_{01}$, 
Alice declares $b_i=0$ (thus winning) if she obtains $M_{00}$ and
$b_i=1$ (thus losing) if she obtains $M_{01}$. 
If Alice wants to declare $b_i=1$, on the other hand, then she
performs a POV measurement $M_1$. She declares $b_i=0$ (losing) if she
obtains $M_{10}$ and $b_i=1$ (winning) if she obtains $M_{11}$. 

What advantage does this strategy give Alice? Suppose that Bob's
(normalized) reduced density matrices, conditioned on Alice getting
the outcomes $M_{00}$, $M_{01}$, $M_{10}$ and $M_{11}$, are $\sigma$,
$\bar{\sigma}$, $\bar{\tau}$ and $\tau$ respectively. Then we can
write 
\begin{eqnarray}
\rho_B&=&q\,\sigma+(1-q)\,\bar{\sigma},\label{m0constraint}\\
\rho_B&=&q'\,\tau+(1-q')\,\bar{\tau},\label{m1constraint}
\end{eqnarray}
where $q$ is the probability of Alice getting outcome $M_{00}$, given
that she performs measurement $M_0$, and $q'$ is the probability of
her getting outcome $M_{11}$, given that she performs measurement
$M_1$. 
It can be shown via a symmetry argument that we do not lose generality if we suppose that
\begin{equation}\label{qconstraint}
q=q'.
\end{equation}
We can also write
\begin{eqnarray}
\rho_B&=&\frac12\left(\rho_0+\rho_1\right),\label{rhobconstraint}\\
\rho_0&=&q\,\sigma+(1-q)\,\bar{\tau},\label{rho0constraint}\\
\rho_1&=&q'\,\tau+(1-q')\,\bar{\sigma}\label{rho1constraint}.
\end{eqnarray}
 
The probability of Alice getting the outcome she wants is given by
$q$, so we have that $\epsilon_A=q-1/2$. The problem is now to
maximize $q$ subject to the constraints of
Eqs.~(\ref{m0constraint})-(\ref{rho1constraint}) (and of course the
constraints that $0\leq q\leq 1$ and that
$\sigma,\bar{\sigma},\tau,\bar{\tau}$ are valid normalized density
operators). Note that if we find a solution for valid
$\sigma,\bar{\sigma},\tau,\bar{\tau}$, then the Hughston-Jozsa-Wootters (HJW) theorem \cite{hjw}
ensures that there does indeed exist a strategy of Alice's that
corresponds to this solution. In other words, there is a state
$\ket{\psi}_{AB}$, and measurements $M_0$ and $M_1$, that give rise to
$\sigma,\bar{\sigma},\tau,\bar{\tau}$ when we condition on Alice's
outcomes. 

We have obtained an upper bound on Alice's cheating capacity that
applies for arbitrary quantum states and noise. We write the
fidelity between a general state $\rho$ and a pure state
$\ket{\psi}$ as $F(\rho,\ket{\psi})\equiv
\bra{\psi}\rho\ket{\psi}$. We write the trace distance between
two general states $\rho$ and $\rho'$ as $D(\rho,\rho')\equiv
1/2|\!|\rho-\rho'|\!|$, where
$|\!|A|\!|=\mathrm{Tr}\sqrt{A^{\dagger}A}$. All the results concerning
these quantities used below can be found in Ref.~\cite{nielsenchuang}
(although note that the fidelity is defined slightly differently). 
\begin{theorem}\label{upperbound}
For all uncorrelated strategies of Alice, we have that for large $n$, 
\begin{equation}\label{upperboundequation}
\epsilon_A\leq\frac{\sqrt{2\gamma}}{\sin^2\theta}.
\end{equation}
\end{theorem}
To prove this bound, note that if Bob is not to abort we must have $F(\rho_0,\ket{\phi_0})\geq 1-\gamma$, and that this,
along with Eq.~(\ref{rho0constraint}), gives 
\[
q\,\bra{\phi_0}\sigma\ket{\phi_0}+(1-q)\, 
\bra{\phi_0}\bar{\tau}\ket{\phi_0}\geq\,1-\gamma.
\]  
This in turn implies 
\[
\bra{\phi_0}\sigma\ket{\phi_0}\geq 1-\gamma/q.
\]
Using the fact that $D(\rho,\rho')\leq\sqrt{1-F(\rho,\rho')}$ for
arbitrary states $\rho$ and $\rho'$, this gives us 
\[
D(\sigma,\ket{\phi_0})\leq\sqrt{\gamma/q}.
\]
Similarly, we can derive
$D(\bar{\tau},\ket{\phi_0})\leq\sqrt{\gamma/(1-q)}$,
$D(\bar{\sigma},\ket{\phi_1})\leq\sqrt{\gamma/(1-q)}$, and
$D(\tau,\ket{\phi_1})\leq\sqrt{\gamma/q}$.
We now recall, from
Eqs.~(\ref{m0constraint})-(\ref{qconstraint}), that 
\[
\rho_B=q\,\sigma+(1-q)\,\bar{\sigma}=q\,\tau+(1-q)\,\bar{\tau}.
\]
Combining this with the above, and using
the fact that
$D(\rho,\rho')=\max_P\left|\mathrm{Tr}(P\,\rho)-\mathrm{Tr}(P\,\rho')\right|$,
where the maximum is over all projection operators, we get that
\begin{eqnarray}
&&q\big(\mathrm{Tr}(P\,\ket{\phi_0}\bra{\phi_0})-\sqrt{\gamma/{q}}
\big)+(1-q)\big(\mathrm{Tr}(P\,\ket{\phi_1}
\bra{\phi_1})-\nonumber\\  
&&\sqrt{\gamma/(1-q)}\big)\;\leq\; q\big(\mathrm{Tr}(P\,
\ket{\phi_1}\bra{\phi_1}) +\sqrt{\gamma/q}\big)+\nonumber\\
&&(1-q)\big(\mathrm{Tr}(P\,\ket{\phi_0}\bra{\phi_0})+
\sqrt{\gamma/(1-q)} \big),
\end{eqnarray}
for any projection operator $P$. Setting $P=\ket{\phi_0}\bra{\phi_0}$
then gives Theorem~\ref{upperbound}.

We have also analyzed in detail the simple case in which the quantum
states are qubit states and the channel is a depolarizing channel,
acting as $\rho\longrightarrow{\cal E}(\rho)\;\equiv\;
f\,\rho+(1-f)\,I/2$. In this case, Alice's optimal cheating strategy
can be found explicitly: 
\begin{theorem}\label{theorem}
For the qubit depolarizing channel, if Alice adopts her optimal
uncorrelated cheating strategy, then for large $n$ 
\begin{eqnarray}
\epsilon_A &=& \frac12\left(1-f\sin\theta\right)\ \ \ \mathrm{if} \
f\leq f^*,\nonumber\\ 
\epsilon_A &=&
\frac12\sqrt{\frac{f^2(1-f^2)\cos^2\theta}{1-f^2\cos^2\theta}}\ \ \
\mathrm{if} \ f>f^*, 
\end{eqnarray}
where
$f^*\equiv (\sqrt{1+3\,\cos^2\theta}-\sin\theta)/2\cos^2\theta$.
\end{theorem}
The proof of Theorem~\ref{theorem} is given in the appendix.

We can compare this result with the upper bound above. If we set
$f=1-2\gamma$, then for fixed $\theta$, we find that
$\epsilon_A\rightarrow\sqrt{\gamma}\cot\theta$, as $\gamma\rightarrow
0$. This shows that the $\gamma$ dependence of
Eq.~(\ref{upperboundequation}) is close to optimal.

In conclusion, we have shown that the attainable security in quantum
coin tossing is qualitatively affected by the presence of noise. 
Indeed in the presence of significant noise, tossing a single coin does not give
acceptable security. Rather, in this case one should consider 
protocols for generation of strings of random bits. 
As we explain above, generating a string of random
bits is  a weaker
protocol than tossing a single coin. However in situations where one
needs to toss coins many times in succession (for instance if one
wants to play repeatedly with a quantum casino), then bit-string generation
can be useful.
The importance of bit-string generation is that even in situations where
tossing a single coin is impossible, it will be possible to generate a
string of bits such  that the average
bias of the bits is bounded. 
We have illustrated this by a simple protocol for
which we prove bounds on the average bias 
in the case where
uncorrelated cheating strategies are used. 

Our work is motivated by the present status of quantum
communication. Indeed with present day optical technology, quantum
communication can be performed over short distances (e.g., laboratory
length scales) with minimal noise and absorption. 
In this case, Theorem~\ref{upperbound} above
indicates that quantum $n$-bit string generation, using our protocol,
should be practically possible with good security.
Over longer
distances (kilometers and above), losses in particular are
significant, and our results would need to be generalized.  

\acknowledgments
We would like to thank Harry Buhrman and Hein Roehrig for useful discussions.
We acknowledge financial support from the Communaut\'{e} Fran\c{c}aise de
Belgique under grant
ARC 00/05-251, from the IUAP programme of the Belgian
government under grant V-18, from the EU under project RESQ
(IST-2001-37559).
\vskip10pt

\appendix{\bf Appendix: Proof of Theorem~\ref{theorem}}

\vskip10pt
To prove Theorem~\ref{theorem}, note that without loss of generality we can write
\begin{eqnarray}
\ket{\phi_0}\bra{\phi_0}&=&1/2\left(I+\alpha\sigma_z-\beta\sigma_x\right),\nonumber\\
\ket{\phi_1}\bra{\phi_1}&=&1/2\left(I+\alpha\sigma_z+\beta\sigma_x\right),
\end{eqnarray}
where $\sigma_x$ and $\sigma_z$ are Pauli sigma matrices, $\alpha^2+\beta^2=1$, and we have that $\alpha=\cos\theta$, $\beta=\sin\theta$.
Recall that the channel acts as
\begin{equation}
\rho\longrightarrow{\cal E}(\rho)\equiv f\rho+(1-f)I/2.
\end{equation}
We then begin by writing
\begin{eqnarray}
\rho_0&=&1/2\left(I+\alpha f\sigma_z - \beta f \sigma_x\right),\nonumber\\
\rho_1&=&1/2\left(I+\alpha f\sigma_z + \beta f \sigma_x\right)
\end{eqnarray}
and
\begin{eqnarray}\label{definitionofsandt}
q\sigma&=&1/2\left(q I+s_x\sigma_x+s_y\sigma_y+s_z\sigma_z\right),\nonumber\\
(1-q)\bar{\sigma}&=&1/2\left((1-q)I+\bar{s}_x\sigma_x+\bar{s}_y\sigma_y+\bar{s}_z\sigma_z\right),\nonumber\\
q\tau&=&1/2\left(qI+t_x\sigma_x+t_y\sigma_y+t_z\sigma_z\right),\nonumber\\
(1-q)\bar{\tau}&=&1/2\left((1-q)I+\bar{t}_x\sigma_x+\bar{t}_y\sigma_y+\bar{t}_z\sigma_z\right),\nonumber\\
\end{eqnarray}
where $-1\leq s_x,s_y,s_z\leq1$, and so on. Conditions (\ref{m0constraint})-(\ref{rho1constraint}) imply
\begin{eqnarray}
s_z+\bar{s}_z&=&f\alpha,\ \ \ \ \ \ s_x+\bar{s}_x=0,\nonumber\\
t_z+\bar{t}_z&=&f\alpha,\ \ \ \ \ \ t_x+\bar{t}_x=0,\nonumber\\
s_z+\bar{t}_z&=&f\alpha,\ \ \ \ \ \ s_x+\bar{t}_x=-f\beta,\nonumber\\
\bar{s}_z+t_z&=&f\alpha,\ \ \ \ \ \ \bar{s}_x+t_x=f\beta,
\end{eqnarray} 
while symmetry considerations imply
\begin{equation}
s_y=\bar{s}_y=t_y=\bar{t}_y=0.
\end{equation}
From the positivity of the matrices $\sigma,\bar{\sigma},\tau,\bar{\tau}$, we then have that
\begin{eqnarray}
s_z^2+s_x^2&\leq&q^2,\nonumber\\
(f\alpha-s_z)^2+s_x^2&\leq&(1-q)^2,\nonumber\\
s_z^2+(f\beta+s_x)^2&\leq&q^2,\nonumber\\
(f\alpha-s_z)^2+(f\beta+s_x)^2&\leq&(1-q)^2.
\end{eqnarray}
Our aim is to maximize $q$ with respect to $s_x$ and $s_z$, subject to the various constraints. The HJW theorem \cite{hjw} will ensure that there do exist measurements $M_0$ and $M_1$, such that conditions~(\ref{m0constraint}), (\ref{m1constraint}), and (\ref{definitionofsandt}) are satisfied.
 
By inspection we see that the maximum value of $q$ can be obtained when $s_x=-f\beta/2$. This leaves
\begin{eqnarray}
s_z^2+1/4(f^2\beta^2)&\leq&q^2,\nonumber\\
(f\alpha-s_z)^2+1/4(f^2\beta^2)&\leq&(1-q)^2.
\end{eqnarray}
If we consider the equation derived from each of these inequalities, we see that each represents a hyperbola in the $qs_z$ plane. Geometrical considerations tell us that there are two cases to be considered. In the first case, we need only find a turning point of the second hyperbola, and it is guaranteed to lie above the first hyperbola, so that the first inequality will be satisfied. This occurs if $f^2\alpha^2\leq 1-f|\beta|$ (i.e., $f\leq f^*$). In this case, it is easy to see that we maximize $q$ by setting $s_z=f\alpha$, giving
\begin{eqnarray}
q&=&1-(1/2)(f|\beta|),\nonumber\\
\epsilon_A&=&(1/2)(1-\sin\theta).
\end{eqnarray}
In the second case, the relevant turning point of the second hyperbola lies below the first, implying that this is not a solution that satisfies both inequalities. This occurs if $f^2\alpha^2> 1-f|\beta|$ (i.e., $f>f^*$). In this case, we find the maximum $q$ by finding the intersection of the two hyperbolae, i.e., by considering both inequalities as equalities. A short calculation then gives
\begin{eqnarray}
q&=&\frac12+\frac12\sqrt{\frac{f^2\alpha^2(1-f^2)}{1-f^2\alpha^2}},\nonumber\\
\epsilon_A&=&\frac12\sqrt{\frac{f^2(1-f^2)\cos^2\theta}{1-f^2\cos^2\theta}},
\end{eqnarray}
which is achieved when
\begin{equation}
s_z=\frac12 f\alpha+\frac{q}{f\alpha}-\frac{1}{2f\alpha}.
\end{equation}
Throughout, we have ignored solutions corresponding to unphysical values of the variables. The explicit forms for Alice's measurements $M_0$ and $M_1$ can be calculated from the values for the $s$s and the $t$s, although we have not done this here.


\end{document}